\documentclass[final,3p,times,twocolumn]{elsarticle}
 \biboptions{comma,sort&compress}
\usepackage{here}
 \usepackage{graphicx}
  \usepackage{epsfig}
  \usepackage{amsmath}

\def\nin{\noindent}
\def\beq{\begin{equation}}
\def\eeq{\end{equation}}
\def\bea{\begin{eqnarray}}
\def\eea{\end{eqnarray}}

\journal{Nuc. Phys. (Proc. Suppl.)}

\begin{document}

\begin{frontmatter}



\title{QCD sum rules and thermal properties of Charmonium in the vector channel
}

\author[ronde,stell]{C. A. Dominguez\fnref{nrf}}
  \ead{Cesareo.Dominguez@uct.ac.za}

\author[puc]{M. Loewe\corref{speak}\fnref{conicyt,anillos}}
  \ead{mloewe@fis.puc.cl}

\author[ucn]{J. C. Rojas\fnref{conicyt}}
  \ead{jurojas@ucn.cl}

\author[stell]{Y. Zhang}
  \ead{Yingwen.Zhang@uct.ac.za}

\address[ronde]{Centre for Theoretical Physics and Astrophysics,University of Cape Town, Rondebosch 7700, South Africa}
\address[stell]{Department of Physics, Stellenbosch University,
                   Stellenbosch 7600, South Africa. }
\address[puc]{Facultad de F\'{\i}sica, Pontificia Universidad Cat\'olica de Chile,
                    Casilla 306, Santiago 22, Chile.  }
\address[ucn]{Departamento de F\'{\i}sica, Universidad Cat\'olica del Norte,
                    Casilla 1280, Antofagasta, Chile.}

\cortext[speak]{Speaker}

\fntext[nrf]{Supported by NRF South Africa}
\fntext[conicyt]{Supported by Fondecyt Chile under grant 1095217}
\fntext[anillos]{Supported by Proyecto Anillos ACT119}

\begin{abstract}
\noindent The thermal evolution of the hadronic parameters of
charmonium in the vector channel, i.e. the $J$/$\psi$ resonance
mass, coupling (leptonic decay constant), total width, and continuum
threshold is analyzed in the framework of thermal Hilbert moment QCD
sum rules. The continuum threshold $s_0$, as  in other hadronic
channels, decreases with increasing temperature until the PQCD
threshold $s_0 = 4\, m_Q^2$  is reached at $T\simeq \, 1.22 \, T_c$
($m_Q$ is the charm quark mass) and the $J$/$\psi$ mass is
essentially constant in a wide range of temperatures. The other
hadronic parameters behave in a very different way from those of
light-light and heavy-light quark systems. The total width grows
with temperature up to $T \simeq \, 1.04 \,T_c$ beyond which it
decreases sharply with increasing T. The resonance coupling is also
initially constant beginning  to increase monotonically around $T
\simeq T_c$. This  behavior strongly suggests that the $J$/$\psi$
resonance might survive beyond the critical
temperature for deconfinement, in agreement with lattice QCD results.\\

\end{abstract}


\end{frontmatter}


\section{Introduction}
\nin
We discuss here the thermal evolution of the hadronic parameters of
$J/\psi$ in the vector channel, using thermal QCD Sum Rules
\cite{BOCH}. We refer the reader to the original article \cite{char}
for details. This technique has been used previously in the
light-light and in the heavy-light quark sector \cite{VARIOUS}-
\cite{HL}, with the following emerging picture: (i) For  increasing
temperature, hadronically stable particles develop a non-zero width,
and resonances become broader, diverging at a critical temperature
interpreted as the deconfinement temperature ($T_c$). The thermal
resonance broadening was first proposed in \cite{CAD0}. ii) Above
the resonance region, the continuum threshold in hadronic spectral
functions, i.e. the onset of perturbative QCD (PQCD), decreases
monotonically with increasing temperature. When  $T \rightarrow T_c$
hadrons disappear from the spectrum. (iii) This scenario is also
supported by the behavior of hadronic couplings, or leptonic decay
constants, which approach zero as $T \rightarrow T_c$. Masses, on
the other hand, do not to provide information about deconfinement.

\smallskip
\noindent The thermal behavior of the heavy-heavy quark correlator
should be different from that involving at least one light quark
since: a) In the light-light and heavy-light quark sector, the PQCD
contribution is dominated by the time-like spectral function
(annihilation term), which is relatively unimportant in relation to
the light quark condensate contribution, being the scattering PQCD
spectral function highly suppressed. Instead, for heavy-heavy quark
systems this term becomes increasingly important with increasing
temperature while the annihilation
 term only contributes near threshold; b) The non-perturbative QCD sector in the operator product expansion (OPE)
  of light-light and heavy-light quark correlators is driven  by
   the light quark condensate, responsible for the behavior of the
   continuum threshold since
   $s_0(T)/s_0(0) \simeq \langle\langle \bar{q} q\rangle\rangle/\langle \bar{q} q\rangle$ \cite{CAD2}-\cite{radii}.
    The light quark condensate
   is the order parameter for chiral symmetry restoration. In contrast, for heavy-heavy quark correlators
   the leading power correction in the OPE is that of the gluon condensate, which has a very different
   thermal behavior.
   In this approach, the critical temperature for
deconfinement is a phenomenological parameter  which does  not need
to coincide  with e.g. the critical temperature obtained in lattice
QCD \cite{lattice}. In fact, results from QCD
 sum rules lead to  values of $T_c$ somewhat lower than those from lattice QCD.
In order to compare with other approaches,  we express our results
in terms of  the ratio $T/T_c$.

     We find for  charmonium in the vector channel that the  continuum threshold, $s_0(T)$,
    decreases with increasing
$T$, being driven  by
     the gluon condensate and the PQCD spectral function in the space-like region, until it reaches
     the PQCD threshold $s_0 = 4\, m_Q^2$  at $T\simeq \, 1.22\, T_c$ ($m_Q$ is the charm quark mass).
     Below this value of $s_0$ the sum rules cease to be valid.
     The $J/\psi$ mass remains basically constant as in the light-light or heavy-light systems.
     We have, however,  a very different thermal evolution of the width and the coupling.  Both
      are almost independent of $T$ up to $T \simeq \, 0.8\, T_c$ where the width begins to
       increase substantially, but then above $T \simeq \, 1.04 \, T_c$ it starts to decrease sharply, and
        the coupling  increases also sharply. This
         suggests
          the survival of the $J/\psi$ resonance above the deconfinement
          temperature.

              The PQCD spectral function in the space-like region plays here a very important role.
              Non-relativistic approaches to charmonium at finite temperature would normally
                miss this contribution. In fact, the complex energy plane in the non-relativistic
                 case would only have one cut along the positive real axis, which would correspond
                  to the time-like (annihilation) region of PQCD. The space-like contribution
                   ($q^2 = (\omega^2 - |\bf{q}|^2) \leq 0$) in the form of a cut in the energy
                    plane centered at the origin for $- |\bf{q}| \leq \omega \leq |\bf{q}|$, would
                    not be present in the non-relativistic case.

\section{Hilbert Moment QCD Sum Rules}
\noindent We consider the correlator of the heavy-heavy quark vector
current at finite temperature
\begin{eqnarray}
\Pi_{\mu\nu} (q^{2},T) \!  =  \! -(g_{\mu\nu} q^2 - q_\mu q_\nu) \;
\Pi(q^2,T)\nonumber\\ = i
 \int\; d^{4} \! x \;
e^{i q x} \; \;\theta(x_0)\; <<|[ V_{\mu}(x) \;, \;
V_{\nu}^{\dagger}(0)]|>>  ,
\end{eqnarray}
where $V_\mu(x) = : \bar{Q}(x) \gamma_\mu Q(x):$, and $Q(x)$ is the
heavy (charm) quark field. The matrix element above is understood to
be the Gibbs average in the quark-gluon basis. The imaginary part of
the vector correlator  in PQCD at finite temperature involves two
pieces, one in the time-like region ($q^2 \geq 4 m_Q^2$), $Im \;
\Pi_a(q^2,T)$, which survives at T=0, and one in the space-like
region ($q^2 \leq 0$), $Im \; \Pi_s(q^2,T)$, which vanishes at T=0.
To leading order in PQCD we find

\begin{eqnarray}
&& \hspace{-.8cm}\frac{1}{\pi}\, Im \,\Pi_a(q^2,T) =
 \frac{3}{16 \pi^2}\int_{-v}^{v}dx \;(1-x^2) \nonumber\\&&\left[1 - n_F\left(\frac{|\mathbf{q}| x + \omega}{2 T}\right)
- n_F\left( \frac{|\mathbf{q}| x -\omega}{2 T} \right)\right],
\end{eqnarray}

\noindent where $ v^2 = 1 - 4 m_Q^2/q^2$, $m_Q$ is the heavy quark
mass, $q^2 = \omega^2 - \mathbf{q}^2 \geq 4 m_Q^2$, and  $n_F(z) =
(1+e^z)^{-1}$ is the Fermi thermal function. In the rest frame of
the thermal bath, $|\mathbf{q}| \rightarrow 0$, the above result
reduces to
\begin{eqnarray}
\!\!\!\frac{1}{\pi}\, Im \,\Pi_a(\omega,T) = \frac{v (3 - v^2)}{8
\pi^2}\;
 \left[1 - 2 n_F (\omega/2 T)\right]
\nonumber\\\times\;\theta(\omega - 2 m_Q) \;.
\end{eqnarray}
The quark mass is assumed independent of $T$, a good approximation
for $T <200 \;\mbox{MeV}$ \cite{mQ}. Only the leading order in
 the strong coupling will be considered here.

The PQCD piece in the space-like region demands a careful analysis.
In the complex energy plane, and in the space-like region, the
correlator $\Pi(q^2)$, Eq.(1), has a cut centered at the origin and
extending between $\omega = -|\mathbf{q}|$ and $\omega = |
\mathbf{q}|$. In the rest frame this cut produces a delta function
$\delta(\omega^2)$ in the imaginary part of $\Pi(q^2)$. The result
is
\begin{eqnarray}
\frac{1}{\pi}\, Im \,\Pi_s(\omega,T) = \frac{2}{ \pi^2}\; m_Q^2\;
\delta(\omega^2)\times\nonumber\\\left[   n_F
\left(\frac{m_Q}{T}\right)\;+ \frac{2\, T^2}{m_Q^2} \;\int_{ m_Q/T
}^{\infty} y \,n_F (y)\, dy \;\right] \;.
\end{eqnarray}

The corresponding hadronic representation is parametrized
 in terms of the ground state resonance,the $J$/$\psi$,
followed
 by a continuum given by PQCD after a threshold $s_0 > M^2_{V}$.
 In the zero width approximation, the hadronic spectral function is
\begin{align}
\!\! \frac{1}{\pi}\, Im \,\Pi(s,T)|_{HAD}   =  \frac{1}{\pi} Im
\,\Pi(s,T)|_{RES}\; \theta(s_0 - s) \nonumber\\ \hspace{-1.0cm}+
\frac{1}{\pi} Im \,\Pi(s,T)|_{PQCD} \;\theta(s - s_0) \nonumber \\
= 2 \, f_V^2(T)  \, \delta(s - M_V^2(T)) \,+ \frac{1}{\pi} Im
\Pi(s,T)_{a} \;\theta(s - s_0)\;,
\end{align}
where $s \equiv q^2 = \omega^2 - \mathbf{q}^2$. The leptonic decay
constant is defined as $<0| V_\mu(0) | V(k)> = \sqrt{2}\; M_V \;f_V
\;\epsilon_\mu \; .$

When considering a finite (total) width  the following replacement
will be understood
\begin{align}
\pi \;\delta(s- M_V^2(T)) \rightarrow
 \frac{M_V(T)
\Gamma_V(T)}{(s-M_V^2(T))^2 + M_V^2(T) \Gamma_V^2(T)}\; ,
\end{align}

 The  hadronic scattering term, due to
current scattering off  D-mesons, is negligible \cite{char}. The
correlation function $\Pi(q^2,T)$, Eq.(1), satisfies a once
subtracted
 dispersion relation. To eliminate the subtraction one can use Hilbert moments, i.e.

\begin{eqnarray}
\varphi_N(Q^2,T) \equiv \frac{(-)^{N}}{(N)!}\,
\Bigl(\frac{d}{dQ^2}\Bigr)^{N} \Pi(Q^2,T)\, \nonumber\\
= \frac{1}{\pi} \int_{0}^{\infty} \; \frac{ds}{(s+Q^2)^{N+1}}\,  Im
\,\Pi(s,T)\; ,
\end{eqnarray}

\noindent where $N = 1,2,...$, and $Q^2 \geq 0$ is an external
four-momentum squared, to be considered as
 a free parameter.
Using Cauchy's theorem in the complex s-plane, the Hilbert moments
become Finite Energy QCD sum rules (FESR), i.e.
\begin{equation}
\varphi_N(Q^2, T)|_{RES} =  \varphi_N(Q^2,T)|_{QCD} \;,
\end{equation}
where
\begin{equation}
\varphi_N(Q^2,T)|_{RES} \equiv \frac{1}{\pi}
\int_{0}^{s_0(T)}\frac{ds}{(s+Q^2)^{N+1}}\, Im \,\Pi(s,T)|_{RES} \;,
\end{equation}

\begin{align}
 && \varphi_N(Q^2,T)|_{QCD} \equiv  \frac{1}{\pi}
\int_{4 m_Q^2}^{s_0(T)}\frac{ds}{(s+Q^2)^{N+1}}\, Im \,\Pi_{a}(s,T)
\nonumber \\ &&+ \frac{1}{\pi}
\int_{0}^{\infty}\frac{ds}{(s+Q^2)^{N+1}}\, Im \,\Pi_{s}(s,T) +
\varphi_N(Q^2,T)|_{NP}  \;,
\end{align}
and  $Im \,\Pi(s,T)|_{RES}$ is given by the first term in Eq.(5)
modified
 in finite-width according to Eq.(6), and the PQCD spectral functions are given by  Eqs.(3) and (4).\\

The dimension d=4 non perturbative term in the OPE is well known in
the literature, see \cite{char} for details. The dependence on N is
quite cumbersome and it is proportional to the gluon condensate
$\left<\left<\frac{\alpha_s}{\pi}G^2\right>\right>$. At low
temperatures, this condensate

has been calculated in chiral perturbation theory \cite{BERN}. In
this framework the condensate remains essentially constant up $ T
\sim T_c \simeq 100 \; \mbox{MeV}$, after which it decreases
sharply. In order to go beyond the low temperature regime of chiral
perturbation theory, lattice QCD provides the right tool. A good
approximation \cite{latticeG} is given by the expression
\begin{equation}
\left<\left<\frac{\alpha_s}{\pi}G^2\right>\right> =
\left<\frac{\alpha_s}{\pi}G^2\right> \left[\theta(T^*-T) +
\frac{1-\frac{T}{T_C^*}}{1-\frac{T^*}{T_C^*}}\theta(T-T^*) \right]
\end{equation}
where $T^*\approx 150$ MeV is the breakpoint temperature where the
condensate begins  to decrease appreciably, and $T_C^*\approx 250$
MeV is the
temperature at which $\left<\left<\frac{\alpha_s}{\pi}G^2\right>\right>_{T_C}=0$.\\
Returning to the $Q^2$ dependence of the Hilbert moments, Eq.(7),
 we shall fix $Q^2$ and $s_0(0)$ from the experimental values
of the mass, the coupling, and the width at T=0. At finite
temperature there are non-diagonal (Lorentz non-invariant)
condensates that might contribute to the OPE.
 Non-gluonic operators are highly suppressed \cite{HL}, \cite{ELE}
 so that they can be safely ignored.
We have considered also a gluonic twist-two term in the OPE
introduced in \cite{GLUON2}, and computed on the lattice in
\cite{MORITA}.
Its impact  is small,(2-6)\%, and plays no
appreciable role in the results.

\section{Results}
\nin

We begin by determining $s_0$ and $Q^2$ at T=0 from the moments,
Eq.(8), and using as input the experimental values \cite{PDG} $M_V =
3.097 \;\mbox{GeV}$, $f_V = 196 \;\mbox{MeV}$, and $\Gamma_V = 93.2
\; \mbox{keV}$, as well as $m_Q = 1.3 \;\mbox{GeV}$, and  \cite{G2}
$ \langle0| \frac{\alpha_s}{12 \pi} G^2 |0\rangle \simeq 5 \times
10^{-3} \mbox{GeV}^4$. In the zero-width approximation one finds
from Eq.(9) that
\begin{equation}
\frac{\varphi_1(Q^2)|_{RES}}{\varphi_2(Q^2)|_{RES}} =
\frac{\varphi_2(Q^2)|_{RES}}{\varphi_3(Q^2)|_{RES}}\;.
\end{equation}

Given the extremely small total width of the $J/\psi$ it turns out
that the above relation also holds with extreme accuracy in finite
width. Using Eq.(8) this leads to
\begin{equation}
\frac{\varphi_1(Q^2)|_{QCD}}{\varphi_2(Q^2)|_{QCD}} =
\frac{\varphi_2(Q^2)|_{QCD}}{\varphi_3(Q^2)|_{QCD}}\;,
\end{equation}
which depends only on the two unknowns $s_0$ and $Q^2$, and provides
the first equation to determine this pair of parameters. The second
equation can be e.g. Eq.(8) with $N=1$. In this way we find that
$s_0 = 11.64 \;\mbox{GeV}^2$, and $Q^2 = 10 \;\mbox{GeV}^2$
reproduce the experimental values of the mass, coupling, and width
of $J/\psi$ within less than 1\%. This whole set of hadronic
parameters will then be used to normalize the corresponding
parameters at finite temperature. In this way, see \cite{char} for
details, we were able to find the thermal evolution of $s_0$, the
$J/\psi$ mass, its width and its coupling. We show here only the
behavior of the  width and the coupling (Figs. 1 and 2) since these
are the most important results of this analysis. \small

\begin{figure}[ht]
\includegraphics[scale=0.6]{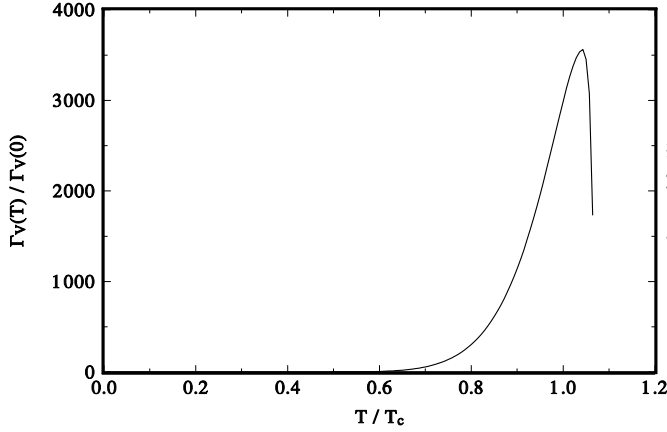}
\caption{The  ratio $\Gamma_V(T)/\Gamma_V(0)$  as a function of
$T/T_c$.}
\end{figure}
\begin{figure}[ht]
\includegraphics[scale=0.6]{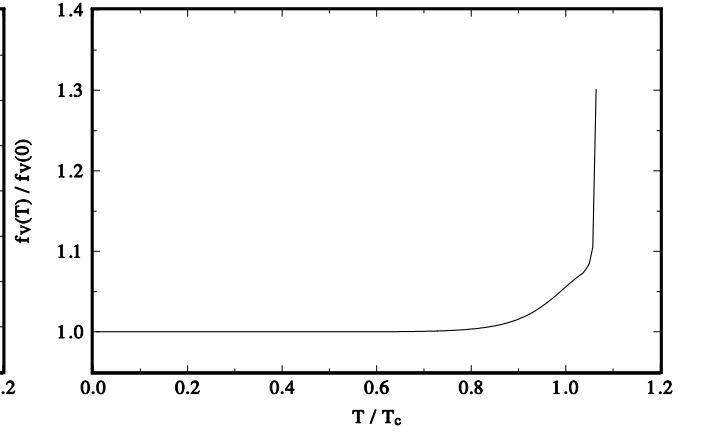}
\caption{The  ratio $f_V(T)/f_V(0)$ as a function of $T/T_c$.}
\end{figure}

\normalsize Both the width and the coupling can only be determined
up to $T_f \simeq\; 1.1 \, T_c$ beyond which $s_0(T) < M_V^2(T)$ and
the FESR integrals have no longer a support. The temperature
behavior of the width and the coupling  shown in Figs. 1 and 2
strongly suggests
the  survival of the $J/\psi$ above the critical temperature for deconfinement.
This conclusion agrees with results from lattice QCD \cite{lattice}, but disagrees
with non-relativistic determinations. As pointed out earlier, the reason for
this disagreement might very well be the absence of the central cut (QCD scattering term)
in the energy plane in non-relativistic frameworks.\\


\nin
%






\begin{thebibliography}{9}



\bibitem{BOCH} A.I. Bochkarev and M.E. Shaposnikov, Nucl. Phys. B 286 (1986) 220.

\bibitem{char} C. A. Dominguez, M. Loewe, J. C. Rojas and Y. Zhang, Phys. Rev. D 81 (2010) 014007.



\bibitem{VARIOUS} R.J. Furnstahl, T. Hatsuda and S.H. Lee,  Phys. Rev. D 42 (1990) 1744;
C. Adami, T. Hatsuda and I. Zahed, Phys. Rev. D 43(1991) 921; C.
Adami and I. Zahed, Phys. Rev. D 45 (1992) 4312; T. Hatsuda, Y.
Koike and S.-H. Lee, Phys. Rev. D 47 (1993) 1225; {\it ibid.} Nucl.
Phys. B 394 (1993) 221; Y. Koike, Phys. Rev. D 48 (1993) 2313.

\bibitem{CAD2} C.A. Dominguez and M. Loewe, Z. Phys. C (Particles \& Fields) 51 (1991) 69; {\it ibid.} 58 (1993) 273;
Phys. Lett. B 481 (2000) 295.

\bibitem{HL} C.A. Dominguez, M. Loewe and J.C. Rojas, J. High Energy Phys. 0708 (2007) 040;
 E.V. Veliev and G. Kaya, arXiV:0902.3443.

\bibitem{CAD0} C.A. Dominguez and M. Loewe, Z. Phys. C (Particles \& Fields) 49 (1991) 423.

\bibitem{radii} C.A. Dominguez, M. Loewe and J.S. Rozowsky, Phys. Lett. B 335 (1994) 506;
C.A. Dominguez, M. S. Fetea and M. Loewe, Phys. Lett.  B 387 (1996)
151; {\it ibid} B 406 (1997) 149; C.A. Dominguez, M. Loewe and  C.
van Gend, Phys. Lett.  B 429 (1998) 64; {\it ibid} B 460 (1999) 442.

\bibitem{lattice} For recent results see e.g. H. Ohno {\it et al.}, arXiv:08103066 and references therein.

\bibitem{mQ} T. Altherr and D. Seibert, Phys. Rev. C 49 (1994) 1684.

\bibitem{BERN} P. Gerber and H. Leutwyler, Nucl. Phys. B 321 (1989) 387.

\bibitem{latticeG} G. Boyd and D. E. Miller, arXiv:hep-ph/9608482 (unpublished);
D.E. Miller, arXiv:hep-ph/0008031 (unpublished)


\bibitem{ELE} V.L. Eletsky, Phys. Lett. B 352 (1995) 440.

\bibitem{GLUON2} F. Klingl, S. Kim, S.H. Lee, P. Morath, and W. Weise, Phys. Rev. Lett. 82, 3396 (1999).

\bibitem{MORITA} K. Morita and S.H. Lee, Phys. Rev. Lett. 100 (2008) 022301; arXiv:0711.3998.

\bibitem{PDG}  Particle Data Group, C. Amsler {\it et al.}, Phys. Lett. B 667, 1 (2008).

\bibitem{G2}   R.A. Bertlmann, {\it et al.}, Z. Phys. C (Particles \& Fields)  39 (1988) 231;
C.A. Dominguez and J. Sola, Z. Phys. C (Particles \& Fields) 40
(1988) 63; C.A. Dominguez and K. Schilcher, J. High Energy Phys. 01
(2007) 093.




\end{thebibliography}








\end{document}